\documentclass[3p,times,procedia,longtitle]{elsarticle}

\usepackage{ecrc}

\volume{00}

\firstpage{1}

\journalname{Physics Procedia}

\runauth{The {\sc Majorana} Collaboration}

\jid{procs}

\jnltitlelogo{Physics Procedia}

\CopyrightLine{2011}{Published by Elsevier Ltd.}

\usepackage{amssymb}
\usepackage{amsmath}
\usepackage{subcaption}

\begin{document}

\def\nuc#1#2{${}^{#1}$#2}
\def\BBz{0$\nu\beta\beta$}
\def\BBt{2$\nu\beta\beta$}
\def\BB{$\beta\beta$}
\def\Tz{$T^{0\nu}_{1/2}$}
\def\Tt{$T^{2\nu}_{1/2}$}
\def\mj{M{\sc ajo\-ra\-na}}
\def\dem{D{\sc e\-mon\-strat\-or}}
\def\mg{M{\sc a}G{\sc e}}
\def\QBB{Q$_{\beta\beta}$}
\def\mBB{$\left < \mbox{m}_{\beta\beta} \right >$}
\def\ge{$^{76}$Ge}

\begin{frontmatter}

\dochead{13th International Conference on Topics in Astroparticle and Underground Physics}

\title{A Dark Matter Search with MALBEK}

\author[unc,tunl]{G.K.~Giovanetti}  
\author[lbnl]{N.~Abgrall}		
\author[pnnl]{E.~Aguayo}
\author[usc,ornl]{F.T.~Avignone~III}
\author[ITEP]{A.S.~Barabash}	
\author[ornl]{F.E.~Bertrand}
\author[lanl]{M.~Boswell} 
\author[JINR]{V.~Brudanin}
\author[duke,tunl]{M.~Busch}	
\author[usd]{D.~Byram} 
\author[sdsmt]{A.S.~Caldwell}
\author[lbnl]{Y-D.~Chan}
\author[sdsmt]{C.D.~Christofferson} 
\author[ncsu,tunl]{D.C.~Combs}  
\author[uw]{C. Cuesta}	
\author[uw]{J.A.~Detwiler}	
\author[uw]{P.J.~Doe} 
\author[ut]{Yu.~Efremenko}
\author[JINR]{V.~Egorov}
\author[ou]{H.~Ejiri}
\author[lanl]{S.R.~Elliott}
\author[pnnl]{J.E.~Fast}
\author[unc,tunl]{P.~Finnerty}  
\author[unc,tunl]{F.M.~Fraenkle} 
\author[ornl]{A.~Galindo-Uribarri}	
\author[lanl]{J. Goett}	
\author[ornl]{M.P.~Green}  
\author[uw]{J. Gruszko}		
\author[usc]{V.E.~Guiseppe}	
\author[JINR]{K.~Gusev}
\author[alberta]{A.L.~Hallin}
\author[ou]{R.~Hazama}
\author[lbnl]{A.~Hegai\fnref{TU}} 
\author[unc,tunl]{R.~Henning}
\author[pnnl]{E.W.~Hoppe}
\author[sdsmt]{S. Howard}  
\author[unc,tunl]{M.A.~Howe}
\author[blhill]{K.J.~Keeter}
\author[ttu]{M.F.~Kidd}	
\author[JINR]{O.~Kochetov}
\author[ITEP]{S.I.~Konovalov}
\author[pnnl]{R.T.~Kouzes}
\author[pnnl]{B.D.~LaFerriere}   
\author[uw]{J.~Leon}	
\author[ncsu,tunl]{L.E.~Leviner}
\author[sjtu]{J.C.~Loach}	
\author[unc,tunl]{J.~MacMullin}
\author[unc,tunl]{S.~MacMullin} 
\author[usd]{R.D.~Martin}
\author[unc,tunl]{S. Meijer}	
\author[lbnl]{S.~Mertens}		
\author[ou]{M.~Nomachi}
\author[pnnl]{J.L.~Orrell}
\author[unc,tunl]{C. O'Shaughnessy}	
\author[pnnl]{N.R.~Overman}  
\author[ncsu,tunl]{D.G.~Phillips~II}  
\author[lbnl]{A.W.P.~Poon}
\author[usd]{K.~Pushkin} 
\author[ornl]{D.C.~Radford}
\author[unc,tunl]{J.~Rager}	
\author[lanl]{K.~Rielage}
\author[uw]{R.G.H.~Robertson}
\author[ut,ornl]{E.~Romero-Romero} 
\author[lanl]{M.C.~Ronquest}	
\author[uw]{A.G.~Schubert}		
\author[unc,tunl]{B.~Shanks}	
\author[ou]{T.~Shima}
\author[JINR]{M.~Shirchenko}
\author[unc,tunl]{K.J.~Snavely}	
\author[usd]{N.~Snyder}	
\author[sdsmt]{A.M.~Suriano} 
\author[blhill,sdsmt]{J.~Thompson} 
\author[JINR]{V.~Timkin}
\author[duke,tunl]{W.~Tornow}
\author[unc,tunl]{J.E.~Trimble}
\author[ornl]{R.L.~Varner}  
\author[ut]{S.~Vasilyev}
\author[lbnl]{K.~Vetter\fnref{ucb}}
\author[unc,tunl]{K.~Vorren} 
\author[ornl]{B.R.~White}	
\author[unc,tunl,ornl]{J.F.~Wilkerson}    
\author[usc]{C.~Wiseman}		
\author[lanl]{W.~Xu}  
\author[JINR]{E.~Yakushev}
\author[ncsu,tunl]{A.R.~Young}
\author[ornl]{C.-H.~Yu}
\author[ITEP]{V.~Yumatov}

\address[unc]{Department of Physics and Astronomy, University of North Carolina, Chapel Hill, NC, USA}
\address[tunl]{Triangle Universities Nuclear Laboratory, Durham, NC, USA}
\address[lbnl]{Nuclear Science Division, Lawrence Berkeley National Laboratory, Berkeley, CA, USA}
\address[pnnl]{Pacific Northwest National Laboratory, Richland, WA, USA}
\address[usc]{Department of Physics and Astronomy, University of South Carolina, Columbia, SC, USA}
\address[ornl]{Oak Ridge National Laboratory, Oak Ridge, TN, USA}
\address[ITEP]{Institute for Theoretical and Experimental Physics, Moscow, Russia}
\address[lanl]{Los Alamos National Laboratory, Los Alamos, NM, USA}
\address[JINR]{Joint Institute for Nuclear Research, Dubna, Russia}
\address[duke]{Department of Physics, Duke University, Durham, NC, USA}
\address[usd]{Department of Physics, University of South Dakota, Vermillion, SD, USA} 
\address[sdsmt]{South Dakota School of Mines and Technology, Rapid City, SD, USA}
\address[ncsu]{Department of Physics, North Carolina State University, Raleigh, NC, USA}
\address[uw]{Center for Experimental Nuclear Physics and Astrophysics, and Department of Physics, University of Washington, Seattle, WA, USA}
\address[ut]{Department of Physics and Astronomy, University of Tennessee, Knoxville, TN, USA}
\address[ou]{Research Center for Nuclear Physics and Department of Physics, Osaka University, Ibaraki, Osaka, Japan}
\address[alberta]{Centre for Particle Physics, University of Alberta, Edmonton, AB, Canada}
\address[blhill]{Department of Physics, Black Hills State University, Spearfish, SD, USA} 
\address[ttu]{Tennessee Tech University, Cookeville, TN, USA}
\address[sjtu]{Shanghai Jiao Tong University, Shanghai, China}
\fntext[TU]{Permanent address: Tuebingen University, Tuebingen, Germany}
\fntext[ucb]{Alternate Address: Department of Nuclear Engineering, University of California, Berkeley, CA, USA}

\begin{abstract}
The {\mj} {\dem} is an array of natural and enriched high purity germanium detectors that will search for the neutrinoless double-beta decay of $^{76}$Ge and perform a search for weakly interacting massive particles (WIMPs) with masses below 10 GeV.  As part of the {\mj} research and development efforts, we have deployed a modified, low-background broad energy germanium detector at the Kimballton Underground Research Facility.  With its sub-keV energy threshold, this detector is sensitive to potential non-Standard Model physics, including interactions with WIMPs. We discuss the backgrounds present in the WIMP region of interest and explore the impact of slow surface event contamination when searching for a WIMP signal.\end{abstract}



\end{frontmatter}

\section{The MALBEK Detector}
The {\mj} collaboration is pursuing the construction of a tonne-scale $^{76}$Ge-based neutrinoless double-beta decay ({\BBz}) experiment~\cite{abg14}. The collaboration's current focus is the assembly of the {\mj} {\dem}, a roughly 40~kg array of of high purity germanium (HPGe) detectors with 30~kg of detectors made from material enriched to 87\% in$^{76}$Ge.  The main goal of the {\dem} is to demonstrate background levels of approximately 3 cnts/(t-y) in the {\BBz} region of interest, low enough to justify the construction of a tonne-scale {\BBz} experiment. The {\dem} uses p-type point contact (PPC) HPGe detectors. These detectors exhibit extremely low electronic noise, allow for pulse shape based background rejection, and are scalable to large arrays. The technical promise of these detectors for physics other than {\BBz}, such as coherent neutrino-nuclear scattering and direct dark matter searches, is presented in~\cite{bar07}. As a part of the research and development efforts for the {\dem}, a customized 465~g Broad Energy Ge (BEGe) PPC detector manufactured by Canberra Industries was installed at the Kimballton Underground Research Facility (KURF) in Ripplemead, Virginia~\cite{fin11}.  This detector is referred to as the {\mj} Low-Background BEGe at Kimballton (MALBEK)~\cite{aal10}. MALBEK's goals were to study the performance and backgrounds of a PPC detector and perform a search for light ($<10\,\mathrm{GeV}/c^2$) weakly interacting massive particle (WIMP) dark matter. Results from an analysis of 89.5~kg-d of data are presented here.

MALBEK was moved to KURF on 12 January 2010. The KURF building is on the 14\textsuperscript{th} level of the Kimballton mine at an overburden of 1450 meters of water equivalent shielding. The mine has drive in access, making transportation of personnel and equipment to the laboratory relatively easy. Power, water, and internet are provided at the site.  The MALBEK detector and shield are housed in the KURF laboratory building within a modified shipping container equipped with humidity controls and an air filtration system. An adjacent shipping container houses the slow control systems and data acquisition (DAQ) electronics. The detector sits inside of a conventional graded shield, comprised of a 2.5~cm layer of ancient lead with $^{210}$Pb activity $<0.01$~Bq~kg$^{-1}$ \cite{fin13} and a 20~cm outer layer of lead purchased from Sullivan Metals. Surrounding the lead shield is a hermetically-sealed acrylic box that is continuously purged with liquid nitrogen boil-off to reduce radon backgrounds. An additional 25~cm layer of polyethylene shields the detector from neutrons.  The thickness of the polyethylene shield was constrained by the internal dimensions of the shipping container that houses the detector.


The MALBEK detector was built by Canberra Industries with custom low-background internal components and an OFHC copper cryostat fabricated at the University of Chicago. MALBEK is outfitted with a low-noise integrated transistor reset preamplifier with two identical signal outputs. One output is AC-coupled directly into a VME-based, Struck Innovative Systeme 3302 (SIS3302) 16-bit digitizer. This unamplified channel has an energy range of $\sim$$0.8-2600$ keV. The second signal output is AC-coupled into a Phillips Scientific 777 fast amplifier before being digitized by the SIS3302. The amplified channel has a range of $\sim$$0.4-150$ keV and is used for the dark matter analysis. The use of a high gain and low gain channel allows for studies of backgrounds over a wide range of energies, from the detector threshold up to the $^{76}$Ge {\BBz} region of interest at 2039~keV. The SIS3302 is running a custom firmware version that implements a trapezoidal filter capable of self-triggering on low amplitude signals.  80~$\mu$s of data are recorded for each trigger to allow for offline event energy reconstruction and pulse shape analysis.  Events that occur in coincidence with data readout from the SIS3302 through the VME backplane are removed in realtime to minimize the effect of noise induced by VME bus traffic. This results in a small, $ <1$\%, decrease in the detector live time. The MALBEK DAQ and slow-control system is controlled by ORCA, an object-oriented data acquisition application with built-in remote monitoring and control capability \cite{how03}. ORCA is used to automate slow control operations, monitor environmental conditions, and provide daily reports to operators on data quality and excursions from standard operating parameters.

\begin{figure}[]
\centering
\begin{subfigure}[t]{0.49\textwidth}
\includegraphics[width=\textwidth]{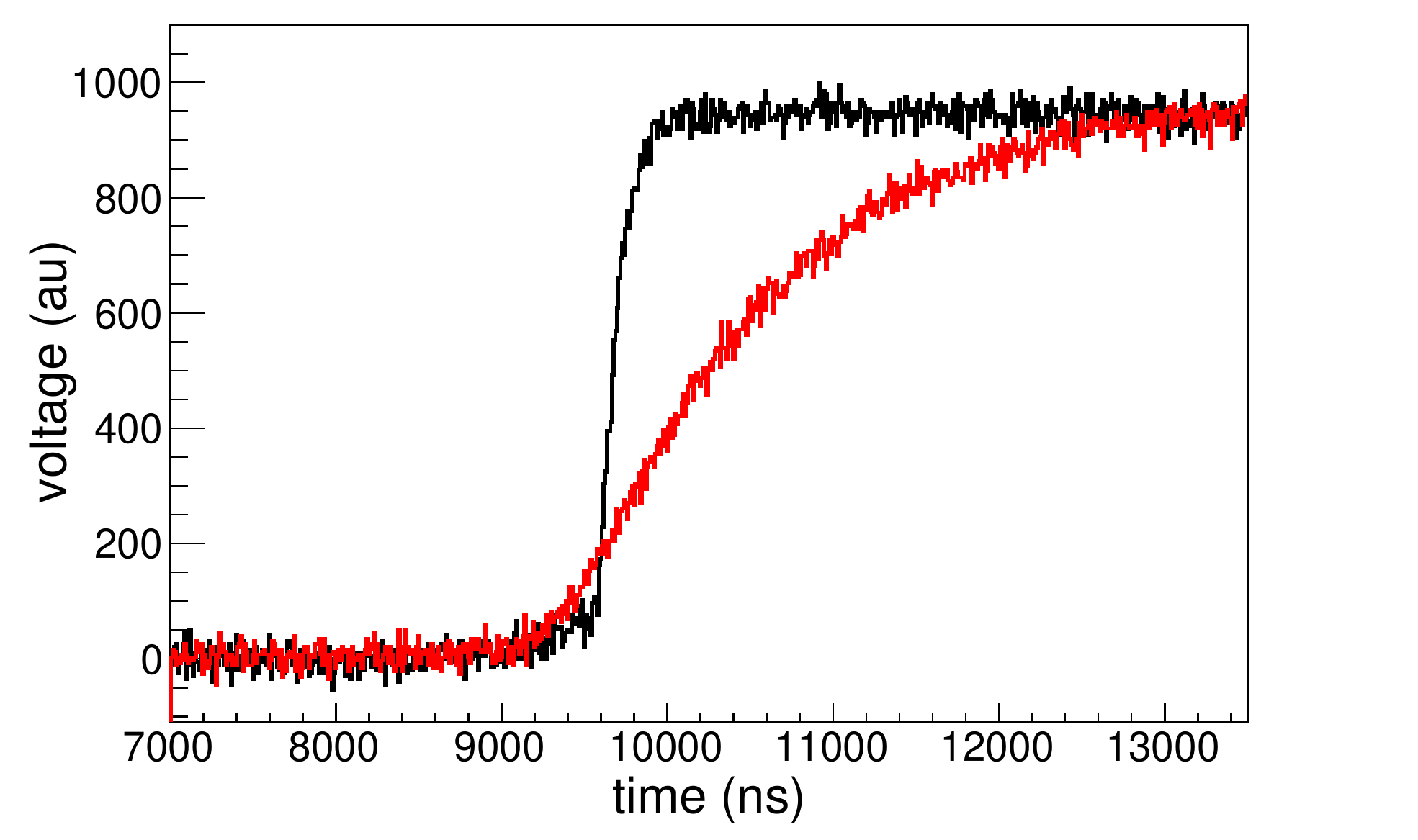}
\caption{}
\label{fig:slow_fast_wf}
\end{subfigure}
\begin{subfigure}[t]{0.49\textwidth}
\includegraphics[width=\textwidth]{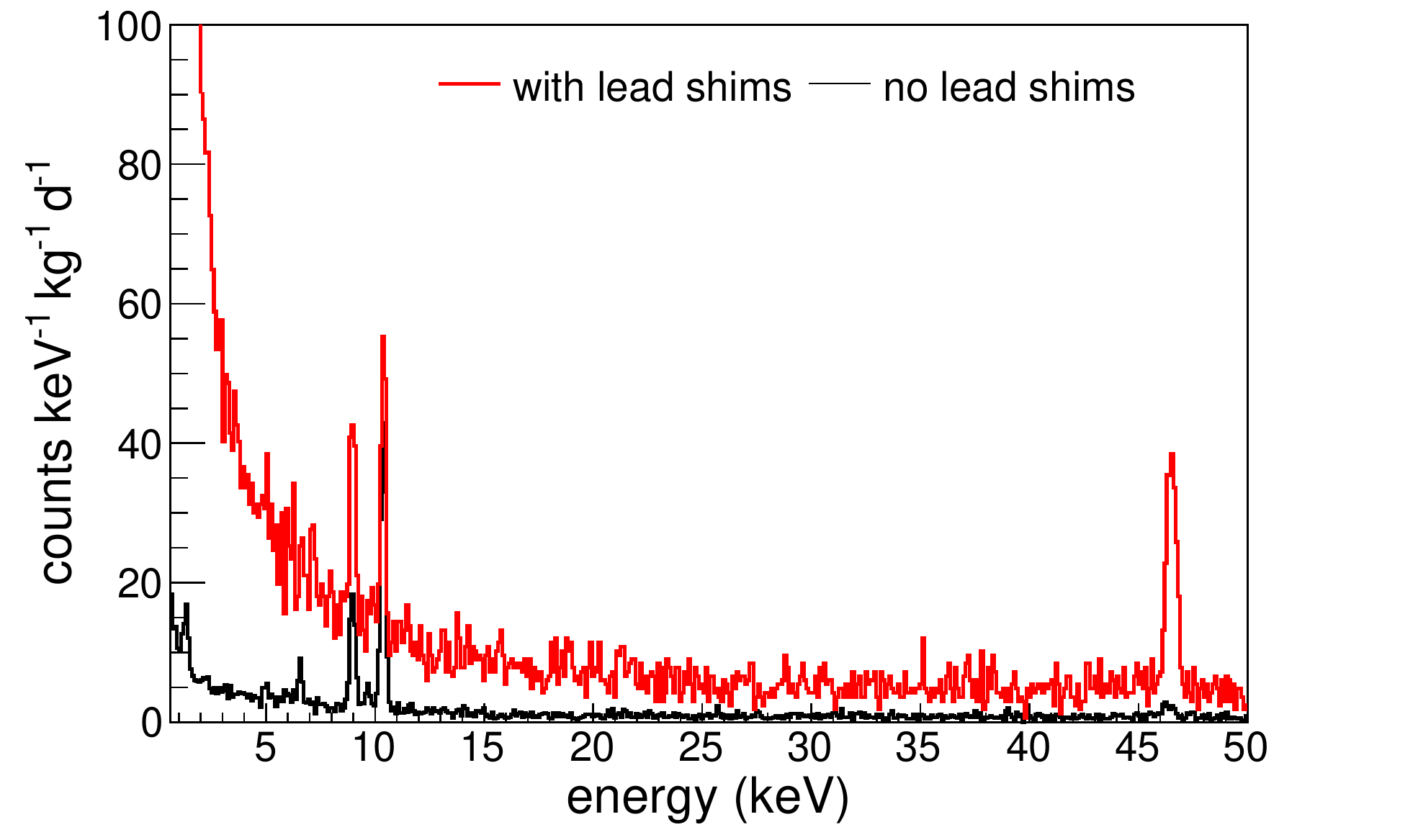}
\caption{}
\label{fig:before_after_pb}
\end{subfigure}
\caption{(a) Two waveforms with energies of approximately 20~keV. The black waveform is a fast rising bulk event. The red waveform is an event that likely occurred near the surface of the detector and has a much longer rise-time. (b) 40 live days of data taken before and after the removal of shims containing high levels of $^{210}$Pb. The 46.5~keV $^{210}$Pb gamma-ray peak is clearly present in the pre-shim removal energy spectrum along with an increased number of events at low energies caused by gamma-ray interactions occurring near the surface of the detector. The post-shim data shows a factor of 10 reduction in the 46.5~keV line and correspondingly fewer events at low energies.}
\end{figure} 

\section{Data Collected with MALBEK}\label{sec:malbek}
The first year of MALBEK operations at KURF were spent testing DAQ configurations for use with the {\mj} {\dem}. In addition to these runs, a series of calibrations were performed with the detector removed from the shield. These calibration measurements will be discussed in detail in an upcoming publication. Two datasets collected with the SIS3302 based DAQ will be described here. The first was collected between 8 March 2011 and 27 May 2011 (79.8~d). These data exhibited a significant and unexpected peak at 46.5~keV originating from $^{210}$Pb. A detailed simulation of the MALBEK internal geometry showed that the likely source of $^{210}$Pb contamination was a pair of lead shims used to hold the Ge crystal within the cryostat ~\cite{sch12}. In October 2011, the detector was driven from KURF to Canberra Industries in Meriden, Connecticut, and the lead shims were replaced with low-background PTFE. The data collected during this period are not suitable for a WIMP search due to the significant background contribution from $^{210}$Pb in the WIMP region of interest. However, the $^{210}$Pb contamination internal to the cryostat provided a low energy gamma-ray calibration source that will be discussed in detail in Section~\ref{sec:slow}. The detector was returned to KURF and the data used for the light WIMP search were acquired between 15 November 2011 and 8 August 2012 (287.9~d).  Due to a period of frequent power outages at KURF, the dataset is divided into two distinct run periods, 15 November 2011 to 12 March 2012 and 9 April 2012 to 29 August 2012.  There were additional periods of dead time within the two run periods caused by intermittent power outages at the mine. The two data periods are calibrated separately using a set of low energy x-ray lines from $^{68,71}$Ge, $^{65}$Zn, and $^{68}$Ga.

During data taking, an arbitrary waveform generator was used to inject $\sim$35 keV signals at a frequency of 0.1~Hz into the test input of the preamplifier.  The location and width of the waveform generator peak in the energy spectrum were measured for every hour long run, enabling the electronic noise and gain drift to be tracked as a function of time. The electronic noise remained stable throughout operation at a level of $164.4\pm0.5$~eV and gain drifts at the detector threshold were smaller than 10~eV. The reset rate of the preamplifier remained constant within 1~Hz during the run period, indicating no periods of increased detector leakage current. An event-by-event timing analysis was done to look for periods of increased electronic noise and it was found that non-poisson processes make up less than 0.5\% of the rate below 1.0~keV.

\section{Slow Events}\label{sec:slow}

A significant background in PPC detectors arises from interactions occurring near the lithium diffused n$^+$ contact on the surface of the detector~\cite{agu13}. Diffusion and recombination processes are hypothesized to dominate charge transport in this region, resulting in energy degraded signals that take longer to reach their maximum amplitude than events originating in the depleted bulk. Representative examples of a slow surface event and a fast bulk event are shown in Figure~\ref{fig:slow_fast_wf}. Figure~\ref{fig:before_after_pb} shows energy spectra collected during 40 live days before and after the lead shims discussed in Section~\ref{sec:malbek} were removed from the MALBEK cryostat. The spectrum collected before the removal of the lead shims shows a peak at 46.5~keV from $^{210}$Pb and a population of events that increase in number towards the detector threshold. These events originate from $^{210}$Pb gamma-rays that, because of their short mean free path in Ge, interact near the surface of the detector and suffer from incomplete charge collection. The spectrum obtained after the removal of the lead shims show a factor of 10 reduction in the magnitude of the 46.5~keV peak and a corresponding reduction in the slow event continuum near the detector threshold. The distribution of the slow rise-time events in the pre-lead shim removal data is roughly exponential with energy, closely mimicking the expected signal from a WIMP interaction in a Ge detector.  It is therefore important to characterize the residual slow event contamination after any cut designed to remove them. 

\begin{figure}
\centering
\begin{subfigure}[t]{0.49\textwidth}
\includegraphics[width=\textwidth]{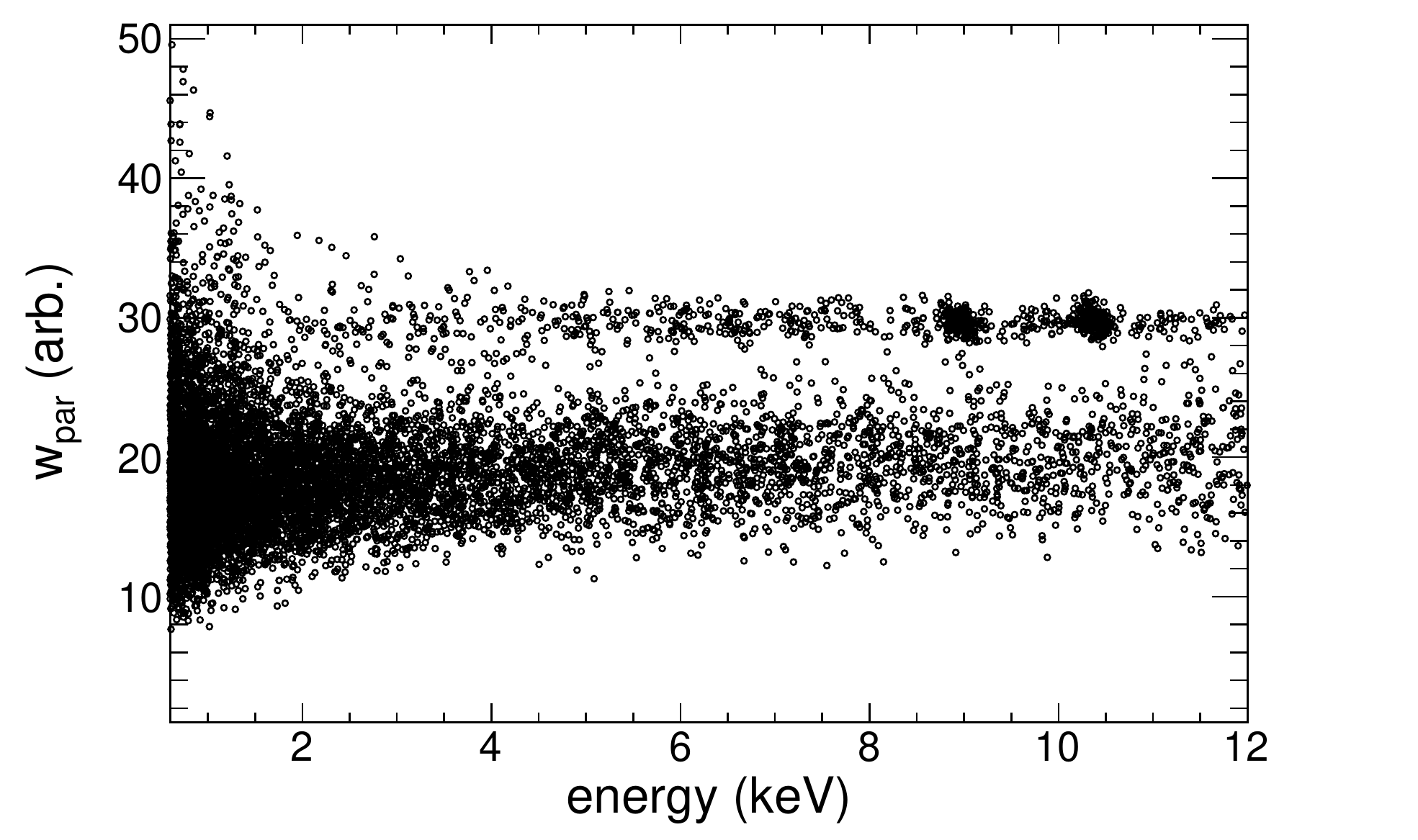}
\caption{}
\label{fig:pb_wpar}
\end{subfigure} 
\begin{subfigure}[t]{0.49\textwidth}
\includegraphics[width=\textwidth]{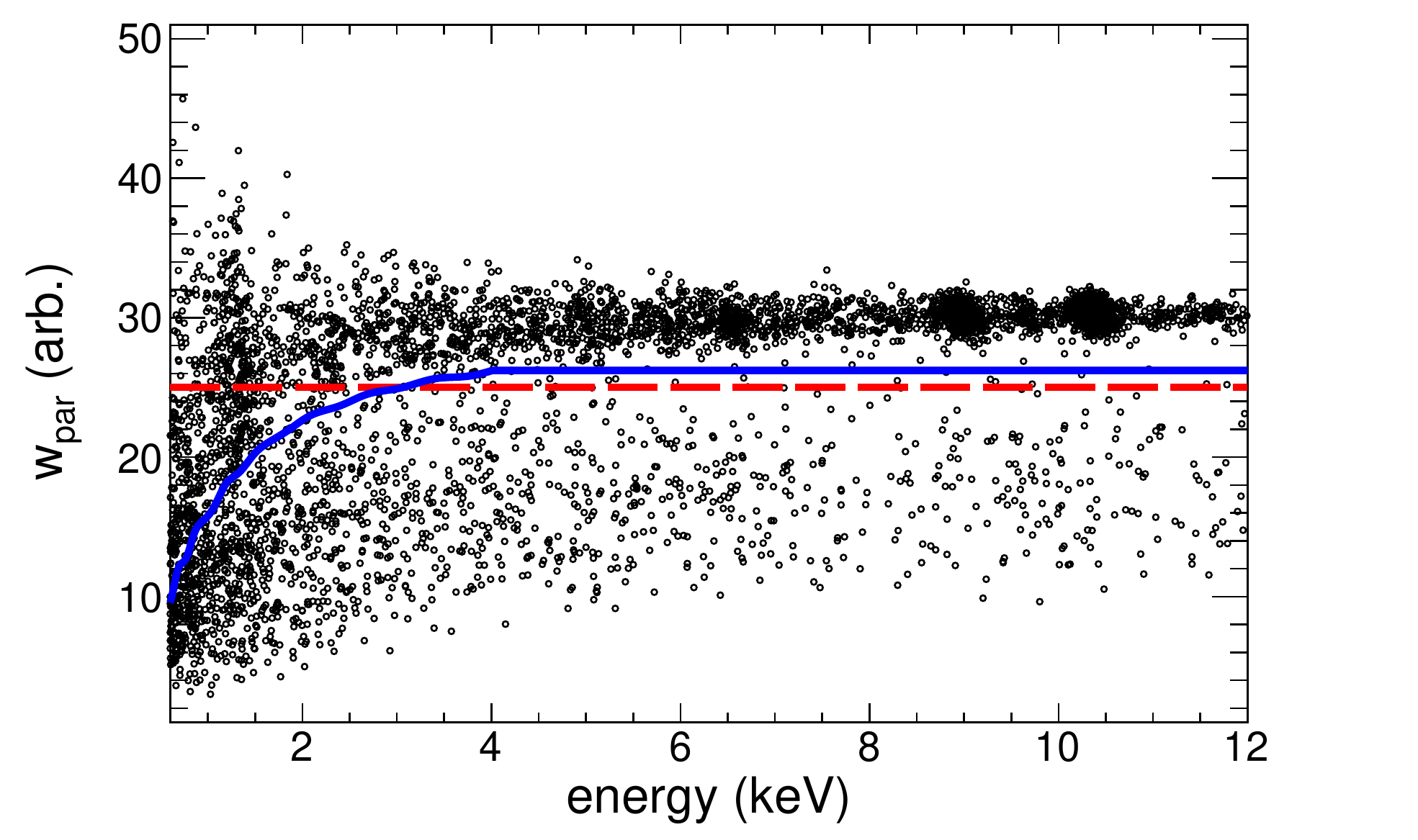}
\caption{}
\label{fig:data_wpar}
\end{subfigure}
\caption{(a) $w_{par}$ distribution of 40 live days of data taken before the lead shims were removed from the cryostat. A band of slow events is centered around $w_{par}=20$. Below 2~keV, this band overlaps with the fast event band centered around $w_{par}=30$.  Events from $^{65}$Zn and $^{68}$Ge x-rays are visible in the fast event band at 8.98~keV and 10.38~keV. (b) $w_{par}$ distribution for 89.5~kg-d of data. The blue line shows the energy dependent slow event cut defined to accept 99\% of fast bulk events. Events below the line are removed. The red line shows the constant slow event cut, designed to remove events in the region where slow pulse contamination is most significant. Again, events below the red line are removed.}
\end{figure}

Slow events are identified by measuring their rise-time, usually defined as the time it takes for a signal to reach some fraction of its maximum amplitude. This can be done by smoothing the waveform to remove noise, calculating the maximum value of the waveform, and scanning along the waveform to determine its rise-time. Applying this technique to MALBEK data with energies above 2~keV results in a clear separation between fast events from interactions that occurred in the detector bulk and slow events from interactions occurring near the surface of the detector. However, below 2~keV, this calculation fails to distinguish slow and fast events due to the reduced signal-to-noise ratio of the digitized waveform. To identify slow events at these energies, a parameter sensitive to rise-time, $w_{par}$, was developed in \cite{fin13} based on the wavelet power spectrum of a digitized waveform. $w_{par}$ is less sensitive to high frequency noise that can be misidentified as maxima during the rise-time calculation. Slower rise-time events have correspondingly smaller $w_{par}$ values. The distribution of $w_{par}$ for 40 live days of data taken before the removal of the lead shims is shown in Figure~\ref{fig:pb_wpar}. The band of events centered around $w_{par}=20$ are surface events with slow rise-times, mostly originating from $^{210}$Pb x-rays. The band of events centered around $w_{par}=30$ are fast bulk events. Peaks from $^{65}$Zn and $^{68}$Ge x-rays can be seen at 8.98~keV and 10.38~keV in the fast event band. 

Despite the improved performance of $w_{par}$ in determining event rise-time, overlap between the slow and fast event bands below 2~keV is evident in Figure~\ref{fig:pb_wpar}. In order to search for a WIMP induced signal within this dataset, one must remove or quantify the slow surface events that contaminate the region of interest. Ideally, this would be done using a predictive model that simulates an event's amplitude and rise-time based on its interaction location within the germanium crystal. A preliminary model is presented in~\cite{fin13} and work towards this goal is ongoing. In the absence of a quantitative description of slow event distributions, an alternative approach must be used. Section~\ref{sec:analysis} describes two methods to address the slow pulse contamination in the WIMP region of interest. The first approach implements a cut on $w_{par}$ that retains 99\% of fast bulk events but accepts an unknown number of slow events. An exponential component is then included in the WIMP spectral fit to model the slow event distribution. The second approach places a stricter cut on $w_{par}$, removing a greater portion of slow events from the dataset at the expense of a reduced efficiency for fast bulk events, including WIMP-nuclear recoils.
 
 \begin{figure}[]
\centering
\begin{subfigure}[t]{0.49\textwidth}
\includegraphics[width=\textwidth]{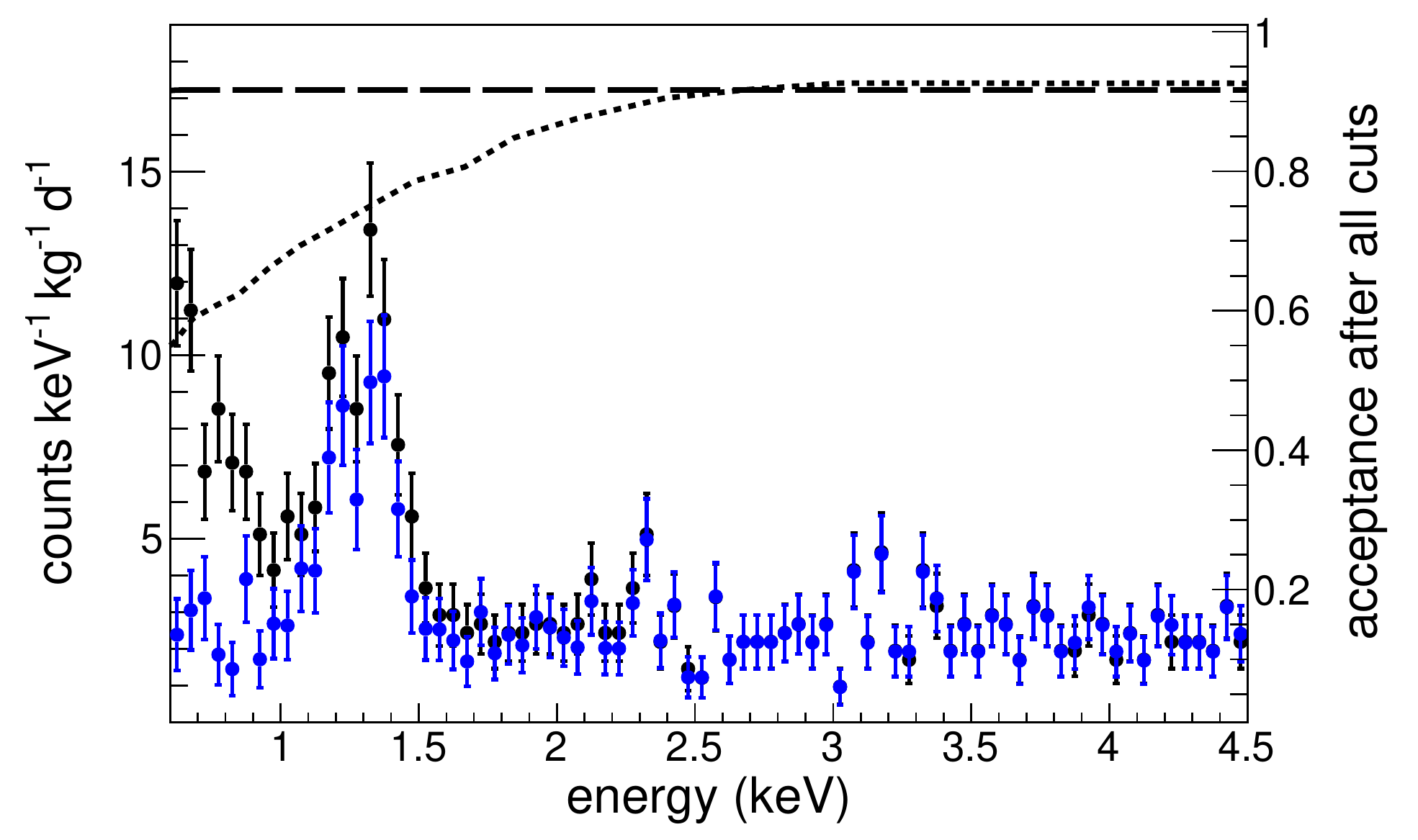}
\caption{}
\label{fig:spectrum}
\end{subfigure} 
\begin{subfigure}[t]{0.49\textwidth}
\includegraphics[width=\textwidth]{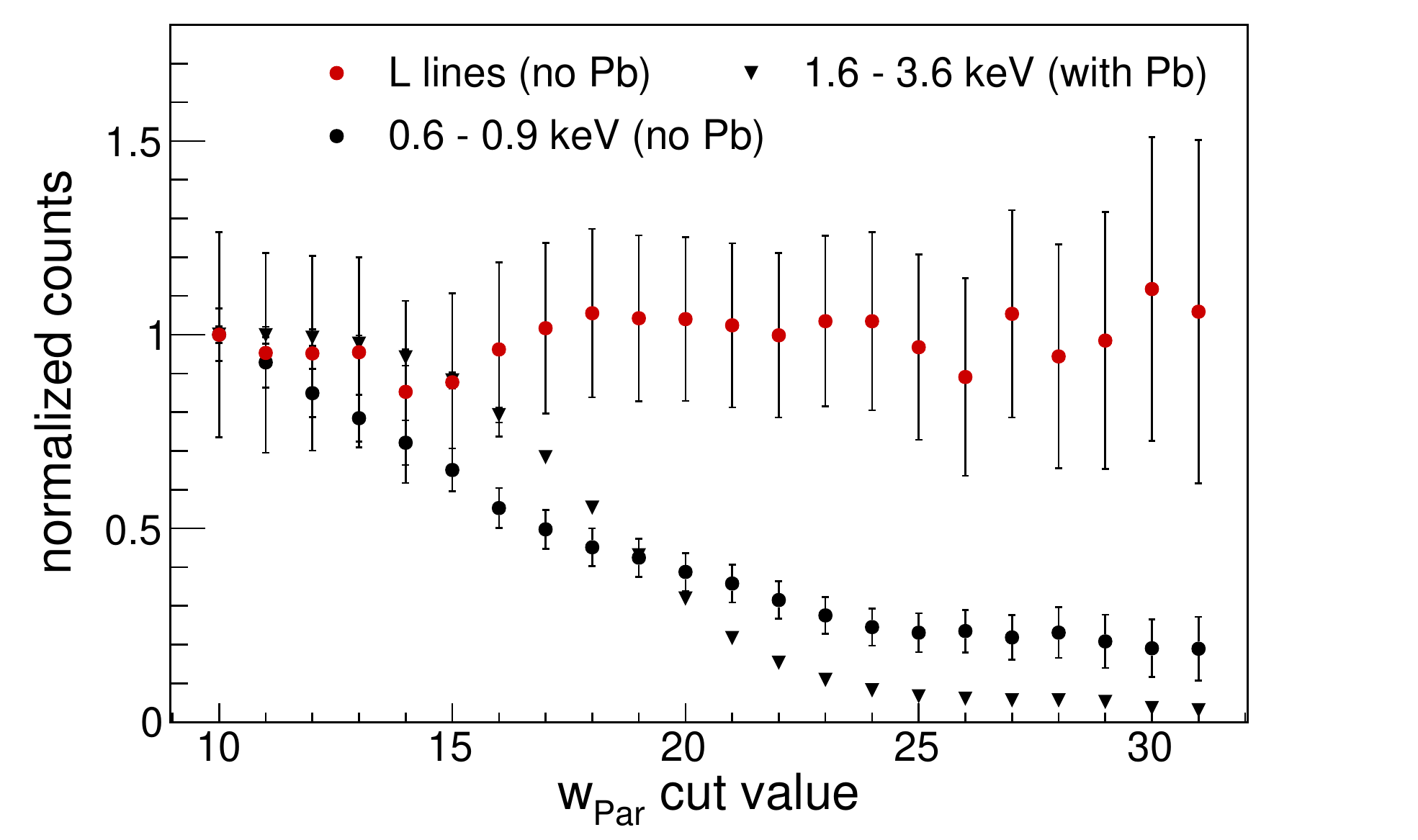}
\caption{}
\label{fig:counts_vs_wpar}
\end{subfigure}
\caption{(a) Energy spectra for the 89.5~kg-d dataset with the 99\% fast acceptance slow event cut (black) and the constant $w_{par}$ cut (blue). The acceptance efficiencies after all cuts are shown for the 99\% acceptance cut (dashed) and the constant cut (dotted). (b) Efficiency corrected counts remaining after a given $w_{par}$ cut. The number of events have been normalized to one so that regions with different rates can be compared in the same figure. The red circles show the normalized counts in the 1.10~keV $^{68}$Ge x-ray peak and the 1.30~keV $^{65}$Zn x-ray peak for the dataset taken after the lead shims were removed. The events contributing to these features occur predominantly in the bulk of the detector and should be unaffected by the $w_{par}$ cut placement. The black circles show the normalized counts in the region between the L peaks and the detector threshold, where significant slow event contamination is expected. The black triangles show the counts in the pre-shim removal data between 1.6~keV and 3.6~keV. The event rate in this region is dominated by slow events originating from $^{210}$Pb gamma-rays interacting near the detector surface.}
\end{figure} 

\section{Data Analysis}\label{sec:analysis}
MALBEK analysis was done using the Germanium Analysis Toolkit (GAT), a modular data analysis framework developed by the {\mj} collaboration. A series of initial cuts were applied to remove non-physics events and periods of high noise from the dataset. Timing cuts were implemented to remove events coincident with preamplifier inhibit pulses, injected test events, and events occurring within 15 minutes of a liquid nitrogen dewar fill. Additional cuts were applied to eliminate non-physics events caused by microphonics and high voltage micro-discharges. Non-timing cuts were designed to accept 99\% of fast bulk events using a dataset of pure fast events generated by injecting pulses with amplitudes ranging from the trigger threshold up to 10~keV into the test input of the preamplifier. Additional details on data selection can be found in \cite{fin13}. The same set of waveform generator induced fast events used to determine the cut efficiencies was used to determine the trigger efficiency of the SIS3302 digitizer. The trigger efficiency drops below 99\% at 550~eV. The analysis threshold was defined to be 600~eV to avoid systematic effects from rapidly changing efficiencies due to small gain shifts at threshold. The live time of the dataset after all cuts is 221.49~d. The active volume of the detector was determined by comparing the ratio of events observed in the 81~keV and 356~keV peaks from a $^{133}$Ba source to a detailed Monte Carlo simulation~\cite{agu13}. It was determined that the full charge collection depth is $933\pm120~\mu\text{m}$.  This reduces the active mass of the detector from $465$~g to $404.2\pm15$~g, resulting in a total exposure of $89.5\pm0.3$~kg-d.
 
\begin{figure}[]
\centering
\includegraphics[width=0.8\textwidth]{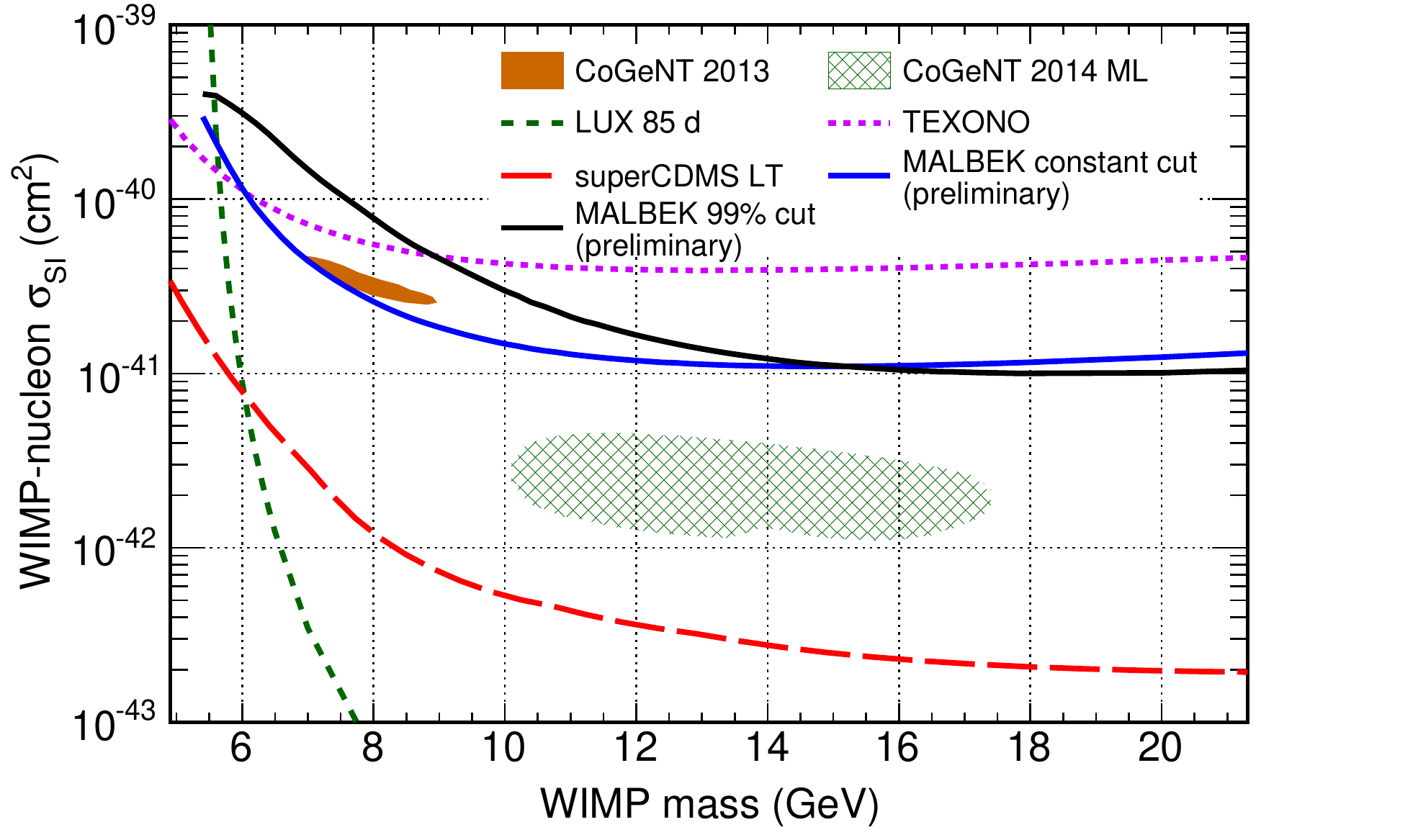}
\caption{90\% upper limits on the spin-independent WIMP-nucleon cross section for the 99\% fast event acceptance cut \emph{(black)} and the constant $w_{par}$ cut \emph{(blue)}. The MALBEK results are shown in comparison to  90\% upper limits from SuperCDMS~\cite{agn14b} \emph{(red long dash)}, LUX~\cite{ake13} \emph{(green dash)}, TEXONO~\cite{li13} \emph{(violet short dash)}, and closed contours from CoGeNT~\cite{aal13, aal14}.}
\label{fig:exclusion}
\end{figure}

Two analyses were performed to search for a WIMP nuclear recoil signal in the 89.5~kg-d dataset. The first analysis utilizes a $w_{par}$ slow event cut that accepts 99\% of the fast pulser events at a given energy, as shown in Figure~\ref{fig:data_wpar}, and treats the unquantified slow event distribution remaining in the signal region as an unknown background as described in \cite{fin13}. The efficiency corrected energy spectrum after all cuts is shown in Figure~\ref{fig:spectrum} along with the calculated fast event acceptance efficiency curve.  An unbinned extended maximum likelihood fit was performed incorporating a flat background component, the $^{65}$Zn and $^{68,71}$Ge L-capture lines, an exponential representing the shape of the unknown slow-signal contamination, and a signal from WIMP induced nuclear recoils.  The WIMP spectrum assumes the expression for the Ge quenching factor described in \cite{aal13} and standard halo assumptions of $v_o=220$~km/s, $v_{esc}=544$~km/s, $\rho=0.3\,\mathrm{GeV}/c^2\,\mathrm{cm}^3$, and the Helm form factor \cite{lew96}. 90\% confidence bounds were determined for WIMP masses ranging from 5.5~GeV to 100~GeV.  The 90\% upper limit on the spin-independent WIMP-nucleon cross section is shown in Figure~\ref{fig:exclusion} along with a selection of recent results from other WIMP searches.

In an effort to reduce the slow event contamination near threshold, a second analysis was performed using a $w_{par}$ cut that is constant with energy as shown in Figure~\ref{fig:data_wpar}.  The constant $w_{par}$ cut removes events from the region where the slow event leakage is the most significant at the expense of a reduced efficiency for accepting fast events. The acceptance efficiency for bulk events is calculated using the fast pulser dataset. The placement of the the constant $w_{par}$ cut was determined by examining the number of efficiency corrected events remaining in different features of the energy spectrum after the cut. Figure~\ref{fig:counts_vs_wpar} shows the efficiency corrected counts in the $^{68}$Ge and $^{65}$Zn x-ray peaks as a function of $w_{par}$ cut value. The counts in the peak are normalized to one to allow for comparison between different regions of the spectrum. The events contributing to these peaks are fast bulk events. If the acceptance efficiency calculated from the fast pulser dataset is correct, the number of efficiency corrected counts in the peaks should remain stable independent of the $w_{par}$ cut placement. Figure~\ref{fig:counts_vs_wpar} also shows the normalized counts in the region $0.6-0.9$~keV, where the slow event contamination is most significant. As the $w_{par}$ cut value increases and more slow events are eliminated, the number of events in the region from $0.6-0.9$~keV decreases. At values greater than $w_{par}=25$, the majority of slow events have been removed. Similar behavior can be seen in the region from $1.6-3.6$~keV in the 40 live day dataset taken with lead shims internal to the cryostat. In this dataset, the region above the $^{68}$Ge and $^{65}$Zn x-ray lines is dominated by slow events originating from the $^{210}$Pb gamma-ray interacting near the surface of the detector. The number of events in this region decreases with decreasing $w_{par}$ until $w_{par}=25$, when the majority of the remaining events are fast bulk events. Figure ~\ref{fig:spectrum} shows the efficiency corrected spectrum with a constant $w_{par}=25$ cut and the bulk event acceptance efficiency curve calculated using the fast waveform generator dataset. 

The insensitivity of the number of efficiency corrected counts in the $^{68}$Ge and $^{65}$Zn x-ray peaks to the $w_{par}$ cut placement, as shown in Figure~\ref{fig:counts_vs_wpar}, provides a check that the fast event efficiency calculated from the waveform generator dataset is correct. At energies below the x-ray peaks, there are no spectral features that allow for a similar test of the calculated fast event acceptance. To account for a possible systematic error in the fast event acceptance efficiency, the 24.7\% uncertainty in the number of counts in the x-ray peaks was used as an estimate of the error in the calculated fast event acceptance across the entire region of interest. The same methods described above were used to search for a WIMP signal, only no exponential component was included in the spectral fit. The 90\% exclusion limits calculated from this analysis are shown in Figure~\ref{fig:exclusion}. 

\section{Conclusions}
The MALBEK detector has been a valuable tool for exploring the backgrounds present in PPC detectors in the WIMP region of interest. Based on this work, there is no evidence for a WIMP induced signal in the 89.5~kg-d of data collected at KURF. It is clear that events that interact near the n$^+$ contact are a significant source of background, and future experiments will need techniques for identifying slow events and models capable of predicting slow event distributions. The {\mj} {\dem} will operate with approximately 100 times the mass of the MALBEK detector and with a lower projected background in the WIMP region of interest. The {\dem} has the opportunity to competitively search for WIMPs with masses below 10~GeV/$c^2$~\cite{mar10} and the {\mj} collaboration is continuing to develop tools that will help accomplish this goal.

\section{Acknowledgments}
The authors would like to thank Juan Collar for providing some of the internal low-background detector components and for useful discussions. We also thank Mike Yocum and Jim Colaresi at Canberra Industries and Sean Finch, S. Derek Roundtree, Werner Tornow, Lhoist North America, and the TUNL technical staff for their logistical support and help with remote operations. We acknowledge support from the Office of Nuclear Physics in the DOE Office of Science, the support of the U.S. Department 
of Energy through the LANL/LDRD Program, the Particle Astrophysics Program of the National Science Foundation, and the Russian Foundation for Basic Research.





\bibliographystyle{elsarticle-num}
\bibliography{gkg_2014_TAUP.bib}

\begin{thebibliography}{10}
\expandafter\ifx\csname url\endcsname\relax
  \def\url#1{\texttt{#1}}\fi
\expandafter\ifx\csname urlprefix\endcsname\relax\def\urlprefix{URL }\fi
\expandafter\ifx\csname href\endcsname\relax
  \def\href#1#2{#2} \def\path#1{#1}\fi

\bibitem{abg14}
N.~Abgrall, et~al., The {\sc majorana demonstrator} neutrinoless double-beta
  decay experiment, Adv. High Energy Phys. 2014.

\bibitem{bar07}
P.~S. Barbeau, J.~I. Collar, O.~Tench, Large-mass ultralow noise germanium
  detectors: performance and applications in neutrino and astroparticle
  physics, JCAP 2007~(09) (2007) 009.

\bibitem{fin11}
P.~Finnerty, S.~MacMullin, H.~O. Back, R.~Henning, A.~Long, K.~T. Macon,
  J.~Strain, R.~M. Lindstrom, R.~B. Vogelaar, Low-background gamma counting at
  the kimballton underground research facility, Nucl. Instrum. Methods A
  642~(1) (2011) 65 -- 69.

\bibitem{aal10}
C.~E. Aalseth, et~al., Astroparticle physics with a customized low-background
  broad energy germanium detector, Nucl. Instrum. Methods A 652~(1) (2011) 692
  -- 695.

\bibitem{fin13}
P.~Finnerty, Ph.D. thesis, University of North Carolina at Chapel Hill (2013).

\bibitem{how03}
M.~A. Howe, G.~A. Cox, P.~J. Harvey, F.~McGirt, K.~Rielage, J.~F. Wilkerson,
  J.~M. Wouters, Sudbury neutrino observatory neutral current detector
  acquisition software overview, IEEE Trans. Nucl. Sci. 1 (2003) 169--173.

\bibitem{sch12}
A.~G. Schubert, Ph.D. thesis, University of Washington (2012).

\bibitem{agu13}
E.~Aguayo, et~al., Characteristics of signals originating near the
  lithium-diffused n+ contact of high purity germanium p-type point contact
  detectors, Nucl. Instrum. Methods A 701 (2013) 176 -- 185.

\bibitem{aal13}
C.~E. Aalseth, et~al., \textsc{C}o\textsc{G}e\textsc{NT}: A search for low-mass
  dark matter using p-type point contact germanium detectors, Phys. Rev. D 88
  (2013) 012002.

\bibitem{lew96}
J.~Lewin, P.~Smith, Review of mathematics, numerical factors, and corrections
  for dark matter experiments based on elastic nuclear recoil, Astroparticle
  Physics 6 (1996) 87 -- 112.

\bibitem{agn14b}
R.~Agnese, et~al., Search for low-mass wimps with \textsc{S}uper\textsc{CDMS},
  arXiv:1402.7137.

\bibitem{ake13}
D.~S. Akerib, et~al., First results from the \textsc{LUX} dark matter
  experiment at the sanford underground research facility, Phys. Rev. Lett. 112
  (2014) 091303.

\bibitem{li13}
H.~Li, et~al., Limits on spin-independent couplings of wimp dark matter with a
  p-type point-contact germanium detector, Phys. Rev. Lett. 110 (2013) 261301.

\bibitem{aal14}
C.~E. Aalseth, et~al., Maximum likelihood signal extraction method applied to
  3.4 years of \textsc{C}o\textsc{G}e\textsc{NT} data, arXiv:1401.6234.

\bibitem{mar10}
M.~G. Marino, Ph.D. thesis, University of Washington (2010).

\end{thebibliography}

\end{document}